%% file: cmelis_iaus289.tex
\documentclass{iaus}
\usepackage{graphicx}
\usepackage{epsfig}
\usepackage{multirow}
\usepackage{natbib}

\title[IAUS 289.~~The Pleiades distance controversy] 
{Toward a VLBI resolution of the Pleiades distance controversy}

\author[C.\ Melis et al.]   
{Carl Melis$^1$, M. J. Reid$^2$, A. J. Mioduszewski$^3$, J. R. Stauffer$^4$, 
 \and G. C. Bower$^5$}

\affiliation{$^1$
Center for Astrophysics and Space Sciences, University of California, San Diego, California 92093-0424, USA;  \\
email: {\tt cmelis@ucsd.edu} \\[\affilskip]
$^2$ Harvard-Smithsonian Center for Astrophysics, 60 Garden Street, Cambridge, MA 02138, USA \\[\affilskip]
$^3$ National Radio Astronomy Observatory, Array Operations Center, 1003 Lopezville Road, Socorro, NM 87801, USA \\[\affilskip]
$^4$ Spitzer Science Center (SSC), 1200 E. California Blvd., California Institute of Technology, Pasadena, CA 91125, USA \\[\affilskip]
$^5$ Astronomy Department \& Radio Astronomy Laboratory, University of California, Berkeley, CA 94720, USA}

\pubyear{2012}
\volume{289}  
\pagerange{1--???}
\jname{Advancing the Physics of Cosmic Distances}
\editors{Richard de Grijs \& Giuseppe Bono, eds.}

\input{astmac}

\begin{document}

\maketitle

\begin{abstract}
The Pleiades is the best studied open cluster in the sky. It is one of the 
primary open clusters used to define the `zero-age main sequence,' and hence it 
serves as a cornerstone for programs which use main-sequence fitting to derive 
distances. This role is called into question by the `Pleiades distance 
controversy' $-$ the distance to the Pleiades from $Hipparcos$ of approximately 120\,pc is 
significantly different from the distance of 133\,pc derived from other 
techniques. To resolve this issue, we plan to use Very Long Baseline Interferometry 
to derive a new, independent trigonometric parallax distance to the Pleiades.  
In these proceedings we present our observational program
and report some preliminary results.
\keywords{techniques: interferometric, astrometry, stars: distances, open clusters and associations: individual (Pleiades), distance scale}
\end{abstract}

\firstsection 
\section{Introduction}

Because of its proximity and its youth, the Pleiades open cluster has been
the subject of extensive observational and theoretical work throughout the
20$^{\rm th}$ century. It remains so in the 21$^{\rm st}$ century, with over 100 refereed
journal papers having `Pleiades' in the title since 2000.
Thanks to the wealth of existing knowledge, the Pleiades cluster
stars are often used as a template with which to define the properties of other
young stars (e.g.\ the Pleiades' lithium abundance vs.\ color is used to define
the locus for PMS stars; the Pleiades' $v$sin$i$ distribution is often compared
to that for other clusters when discussing the evolution of angular momentum on
the main sequence; the first brown dwarf in an open cluster was a Pleiades 
member, and the Pleiades now has the best defined substellar locus of any open
cluster). One would expect that all critical astrophysical parameters for
such an important sample of stars would be well characterized. However, there
still remains an open debate regarding the distance to the Pleiades.

Currently there are two main camps. On one side is the $Hipparcos$ team
\citep{vanleeuwen97,vanleeuwen07}
who state that their satellite's trigonometric parallaxes put the Pleiades
at a distance of 118.3$\pm$3.5\,pc and, more recently, 122.2$\pm$1.9\,pc 
\citep{vanleeuwen97,vanleeuwen07}. On the other side are various ground-based and
$Hubble$ $Space$ $Telescope$ ($HST$)-based teams employing methods from 
main sequence fitting to dynamical
parallax determinations from binary stars (see Table 1).
These teams, whose work can be theory dependent or rely on a small sample of stars, offer a
distance of 133$\pm$0.9\,pc (see Table 1 for a summary of Pleiades distances). 
Outside the $Hipparcos$ community, the most often cited physical mechanism to 
explain the $Hipparcos$ distance to the Pleiades is that there are unmodeled
correlations in the $Hipparcos$ data on angular scales of $\sim$1$^{\circ}$ which
can (under some circumstances) bias distance estimates
\citep{narayanan99}.

Although what is listed above amounts to a 10\% difference in the distance, the 
resultant 
discrepancies as propogated into the Pleiades Hertzsprung-Russell diagram, and the necessary
revisions of physical models to obtain agreement with the Hipparcos result, are
quite significant. The $Hipparcos$ result, if correct, means 
that stars in the Pleiades are on order $\sim$0.2 magnitudes fainter than 
otherwise similar field stars. According to \citet{soderblom05}:

\begin{quotation}
``This large discrepancy has forced a careful reexamination of the assumptions
and input parameters of the stellar models, as well as a 
thorough study of the $Hipparcos$ data itself and potential errors in 
it. The controversy has not been fully resolved in that builders of 
star models find that the changes in physics or input parameters 
needed to account for the $Hipparcos$ distance are too radical to be 
reasonable, whereas the $Hipparcos$ team has resolutely defended 
the $Hipparcos$ result.''
\end{quotation}

The final comment regarding the $Hipparcos$ team has held true despite a recent
`new' reduction of the $Hipparcos$ raw data \citep{vanleeuwen07}. As
stated by van Leeuwen:

\begin{quotation}
The new $Hipparcos$ reduction results largely confirm the earlier results, 
including what has been referred to as errors in the published data: the 
parallaxes of the Pleiades... The new reduction leaves little, if any, room
for an explanation of these differences as due to errors in the $Hipparcos$ data.
\end{quotation}


\begin{table}[t!]
\begin{center}
\caption{Pleiades parallaxes (updated from \citealt{soderblom05})}
\begin{tabular}{lcccc}
\hline \hline
Method & $\pi$$_{\rm abs}$ (mas) & D (pc) & m$-$M & Ref. \\
\hline
$Hipparcos$ all-sky & 8.45$\pm$0.25 & 118.3$\pm$3.5 & 5.37$\pm$0.06 & 2 \\
$Hipparcos$ new reduction & 8.18$\pm$0.13 & 122.2$\pm$1.9 & 5.44$\pm$0.03 & 7 \\
\hline
Main-sequence fitting & 7.58$\pm$0.14 & 131.9$\pm$2.4 & 5.60$\pm$0.04 & 1 \\
Allegheny Observatory parallaxes & 7.64$\pm$0.43 & 130.9$\pm$7.4 & 5.59$\pm$0.11 & 3 \\
Interferometric orbit & 7.41$\pm$0.11 & 135.0$\pm$2.0 & 5.65$\pm$0.03 & 4 \\
Dynamical parallax & 7.58$\pm$0.11 & 131.9$\pm$3.0 & 5.60$\pm$0.05 & 5 \\
$HST$ FGS parallax of 3 Pleiads & 7.43$\pm$0.17 & 134.6$\pm$3.1 & 5.65$\pm$0.05 & 6 \\
\hline
\end{tabular}
\\
References.$-$(1) \citet{pinsonneault98}, (2) \citet{vanleeuwen99}, (3) \citet{gatewood00}, (4) \citet{pan04}, (5) \citet{munari04}, (6) \citet{soderblom05}, (7) \citet{vanleeuwen07}.
\end{center}
\end{table}

What can be done to reach a resolution regarding the distance to the Pleiades?
Do models fall short of describing the $Hipparcos$ Pleiades main sequence due to 
an important,
albeit overlooked, additional physics? \citet{vanleeuwen99} considered whether 
plausible errors in the assumed helium or metal abundance of the Pleiades could 
explain the distance discrepancy, but concluded this seemed very unlikely. The
difference is instead ascribed to some unspecified, age-related property that causes young stars
to be underluminous relative to current theoretical models \citep{vanleeuwen99}. 
Or does $Hipparcos$ contain a systematic or instrumental error that has yet to be 
characterized?
A clear resolution to the Pleiades distance problem requires a new approach
that is free of the limitations of previous optical astrometric measurements.
Such a technique is radio interferometric astrometry as afforded by Very
Long Baseline Interferometry. The highly accurate radio reference frame combined with
the exquisite precision of VLBI astrometric measurements (e.g., \citealt{loinard07}; \citealt{loinard08}; \citealt{reid09}) can be used to settle the Pleiades distance debate.

\section{NRAO Key Science Project}

Using the full High Sensitivity Array (HSA: Very Long Baseline Array, 
Green Bank, Effelsberg, and Arecibo antennas) 
we are conducting a large ($\approx$900~hr) program
to determine the most accurate trigonometric parallax to the Pleaides
cluster and hence resolve the `Pleiades distance controversy'. 

Of course, one needs radio sources with sufficient flux to enable
VLBI measurements. Previous studies of the Pleiades at radio wavelengths have
proven largely unsuccessful (e.g., \citealt{bastian88}, 
\citealt{lim95} and references therein). Pleiades members
have only been detected in a deep survey carried out by \citet{lim95}. 
However, the lesson learned through the study of Lim \& White is that some 
Pleiads have quasi-steady radio luminosities on the order of 
2$\times$10$^{15}$ ergs Hz$^{-1}$ s$^{-1}$. Such luminosities equate to 
flux levels on the order of $\sim$0.2 mJy. Capitalizing on these previous
observations, and with the
eventual goal of a VLBI survey in mind, we attempted deep Very Large Array
(VLA) observations
of the brightest X-ray emitting Pleiads. 
Our VLA sample targeted ultra-fast rotators (UFRs) that have X-ray
luminosities on the order of log(L$_{\rm X}$)$\sim$30 [ergs s$^{-1}$]. 
UFRs are known to exhibit enhanced coronal
activity and are often detectable non-thermal radio emitters.
It is noted that this target selection strategy did not take into account
whether or not sources were suspected members of binary systems.
That is to say, our input VLA target catalog was unbiased in respect to
binarity and included roughly equal numbers of (believed) single and binary
stars. We designed our
VLA experiment to test the quasi-steady flux level of known radio-emitting Pleiads,
aiming for rms flux levels of $\sim$16 $\mu$Jy bm$^{-1}$.
Our program was successful (Table 2), with a $\sim$50\% detection rate when
we reached our sensitivity threshold. The flux levels we measure are on the order
of 50-100 $\mu$Jy.

The HSA is
capable of detecting the elevated flux levels ($\approx$100\,$\mu$Jy) and will be able to
obtain even deeper detections (30-50\,$\mu$Jy) once 2 Gbps sampling,
the phased Karl G.\ Jansky VLA (JVLA), and the VLBA C-band receiver upgrades are complete.

\begin{table}[t!]
\begin{center}
\caption{VLA-detected Pleiads}
\begin{tabular}{lcccccc}
\hline \hline
Star  &  log(L$_{\rm X}$) &   B$-$V &  $v$sin$i$ &  Radio Program  & Flux & Binary? \\
     &  (ergs s$^{-1}$) &  (mag) &      (km s$^{-1}$) &  &   ($\mu$Jy) &  \\
\hline
 HII 174 & 30.19 & 0.81  &  28    &  AM978,JVLA  & 90-120  & Y \\ 
 HII 253 & 30.46 & 0.64  &   37   &  AL361 & 90  & N \\ 
 HII 314 & 30.28 & 0.60 &   38     &  JVLA  &  115 & N \\
 HII 625 & 30.19 & 0.78  &   94   &   LW95,JVLA & 110-160 & Y \\ 
 HII 1136& 30.14 & 0.72  &   75   &   LW95,AL361 & 110-930 & Y \\ 
 HII 1883& 29.67 & 0.99 &140     & LW95 & 50-100 & N \\  
 HII 2147& 30.5    & 0.76  &   27   &   AM978 & 130-180 & Y \\ 
 HII 2244& 29.99  & 0.99  &  45    &  JVLA    &  60 & N \\
 HII 3197& 30.14  & 1.03  &   33  &   AM978 & 90 & Y \\ 
 PELS75 & 30.1   & 0.91   &  56  &   JVLA     & 150 & Y \\
\hline
\end{tabular}
\\
X-ray data taken from \citet{stauffer94} and \citet{micela96,micela99}.
LW95 = \citet{lim95}.
The binary column indicates whether any hint of binarity is noted in literature studies
of the Pleiades (e.g., \citealt{mermilliod92}, \citealt{bouvier97}, and references therein).
\end{center}
\end{table}

\subsection{Project Path}

Our VLBI observational strategy includes nine total epochs of positional measurements
for 10 Pleiads. Five target sources
are being monitored with the HSA and will continue to be monitored until early
2013; after that time five new sources will be monitored for roughly one year. 
This strategy is driven by two considerations: target binarity and
cluster depth issues.

\noindent {\bf Binarity}

\noindent Through astrometric monitoring and literature searches we have determined
that most (if not all) of our 
VLBI target sources reside in binary systems. This preference for binary systems
was not explicit in our source selection procedure; our targets were selected based solely
on bright radio emission detected in preliminary VLA surveys. 
There is mounting evidence that radio-bright stars tend to reside within binary systems,
so it is unlikely that there is a population of radio-loud targets in single systems.

In some special cases, suspicions of binarity
are confirmed by detection of light emitted by the binary component
(e.g., Figure 1). To properly determine the parallax motion of any
astrometric binaries requires
a complete mapping of the binary system's orbital motion. To
do this requires $\approx$9 astrometric measurements spaced over one year.
Such a data set (18 measurements taking R.A.\ and Dec as independent)
will allow us to decouple parallax and proper motion
(5 model parameters) from orbital motion (7 model parameters).
Longer period ($>$1~year) systems, despite
having incomplete orbital period information, may still have an
accurate parallax determined by the inclusion of acceleration terms
in place of complete orbital fits. Such a strategy was successfully
implemented by \citet{loinard07} in their determination of the parallax
of the T Tau binary system with the VLBA.

\noindent  {\bf Cluster Depth}

\noindent The sample size of Pleiades objects
is necessary to decouple cluster-depth issues from individual cluster
member parallax measurements.
In the new reduction of the $Hipparcos$ data, \citet{vanleeuwen07}   
raises an important issue when discussing the $HST$ parallax
measurements of three low-mass cluster members by \citet{soderblom05}.
To derive the cluster absolute parallax, one must include with the
measurements of the individual stars the additional uncertainty of the
star's position with respect to the center of the cluster. 
It is thus of the
utmost importance to have enough members to average out
positionally dependent effects like the (unknown) distance
between the target source and the true cluster center.
A rough estimate of this uncertainty can be made (with simplifications)
as follows: the half mass radius of the Pleiades is 1.9 pc 
\citep{raboud98}. With
10 targets, and if the
uncertainties are dominated by the physical depth of the
cluster, 
our final distance uncertainty would be of order 1.9/$\sqrt{10}$,
or $\sim$0.6 pc.

\section{Preliminary Results}

One star in particular in the Lim \& White sample exhibited a radio flare that peaked at 
a flux density of $\sim$1\,mJy. 
This star, HII\,1136, became the subject of VLBI pilot surveys to determine
the feasibility of a full-scale Pleiades parallax program. We have now amassed
VLBI detections for this system spanning almost 10~years. With these data
we attempt a preliminary parallax fit. Two automated least-squares
fits are performed (for fit details see, e.g., \citealt{reid09} and references therein): 
one fit only allows parallax
and proper motion as free parameters while in the other fit we also allow for a
constant acceleration term, the likes of which would be obtained if the HII\,1136
system was composed of a long orbital period ($>$10~years) binary.

The preliminary fit for the case of a constant acceleration term
is shown in Figure 2. It is noted that the addition of the constant acceleration
term, although resulting in lower rms residuals, does not affect the parallax obtained.
For this particular system we obtain a parallax of $\approx$7.2\,mas. Such a distance is 
slightly farther away than non-Hipparcos distances reported in Table 1. It could be the case 
that HII\,1136 is on the far side of the cluster as it lies within the cluster tidal radius 
regardless of which cluster distance is assumed.

\begin{figure}[!h]
 \centering
 \begin{minipage}[t!]{100mm}
  \includegraphics[width=100mm,angle=-90]{cmelis_fig1a_iaus289.EPS}
  \end{minipage}
  \\
  \begin{minipage}[h!]{44mm}
   \includegraphics[width=45mm,height=45mm]{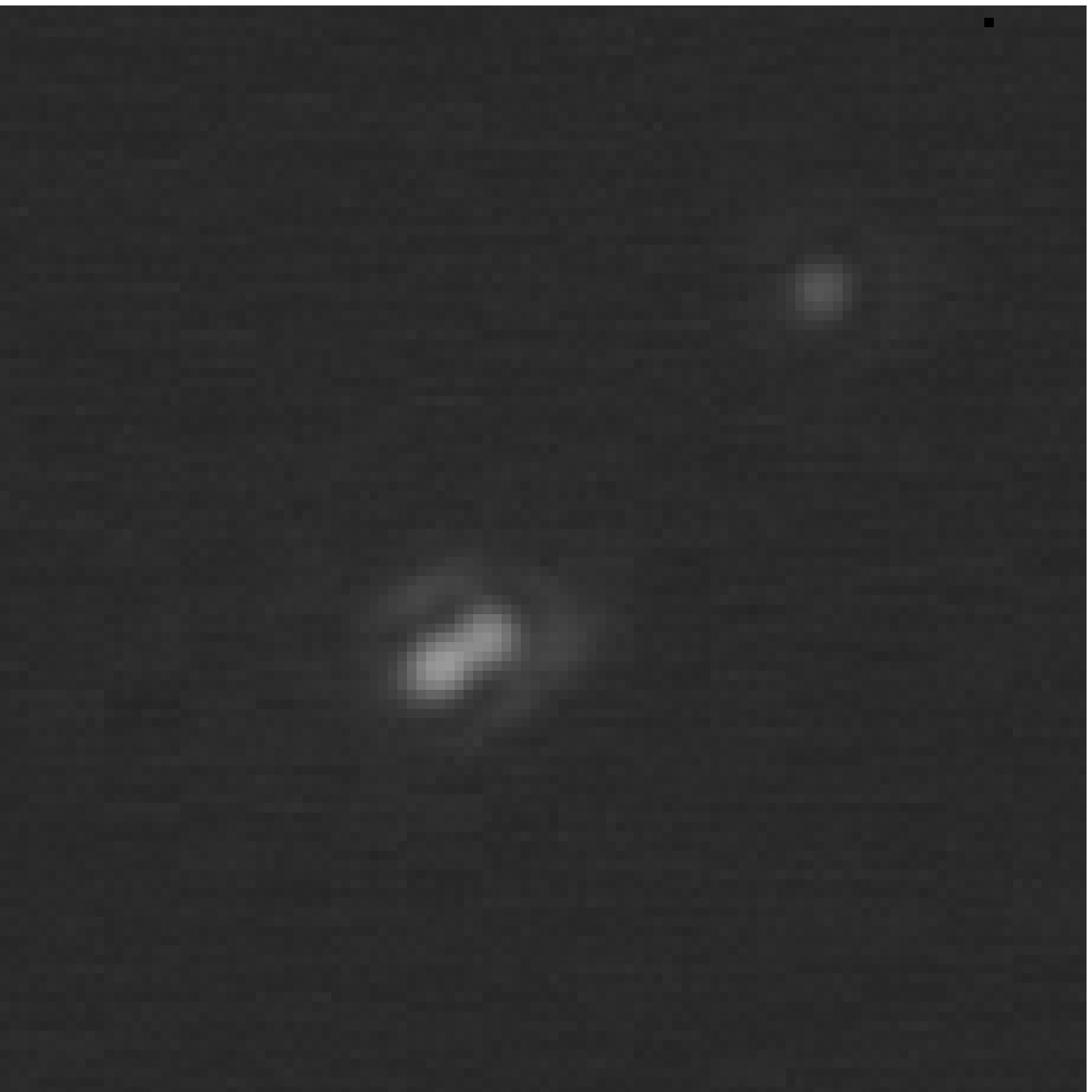}
  \end{minipage}
  \begin{minipage}[h!]{44mm}
   \includegraphics[width=45mm,height=45mm]{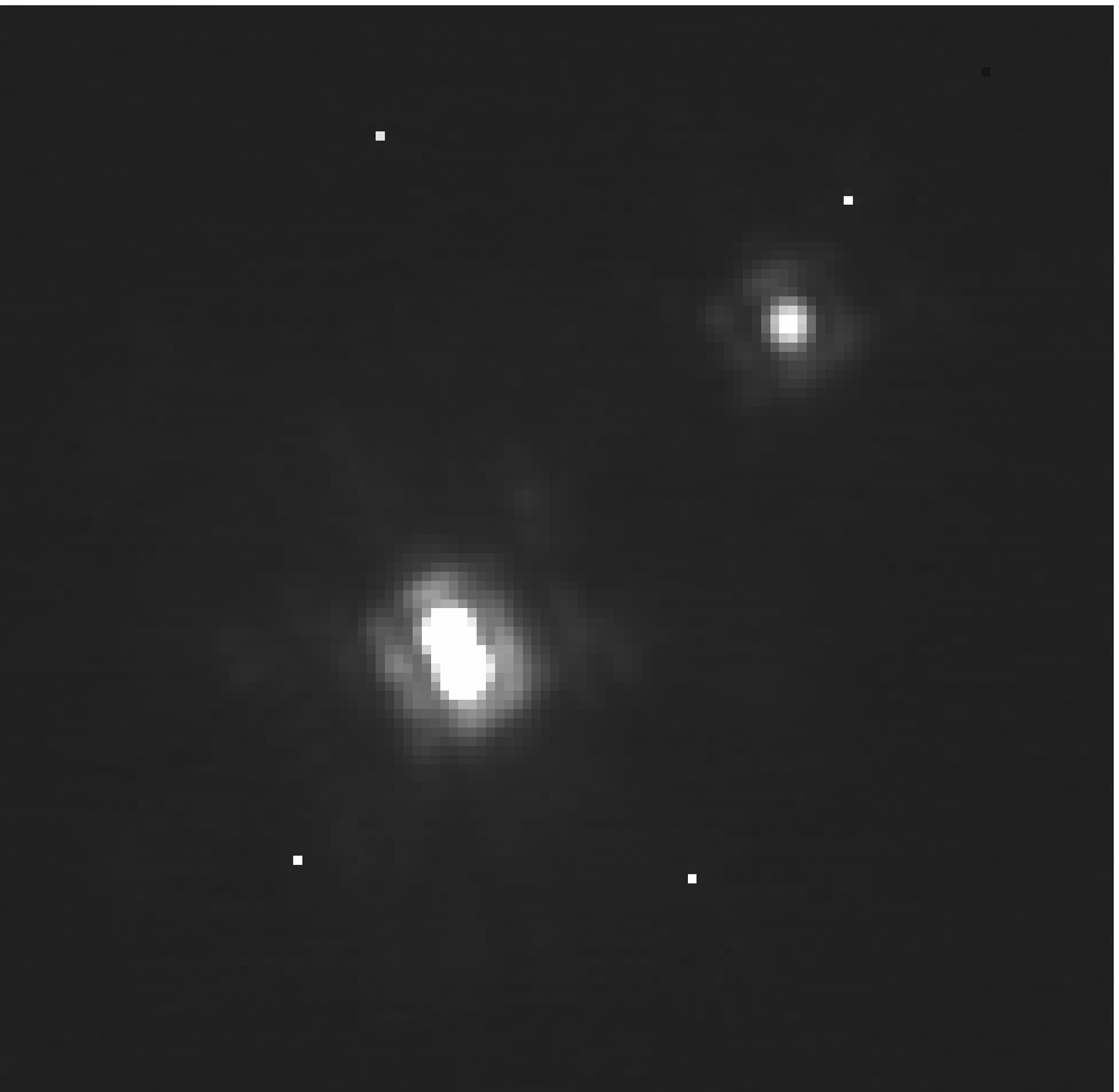}
  \end{minipage}
  \begin{minipage}[h!]{44mm}
   \includegraphics[width=45mm,height=45mm]{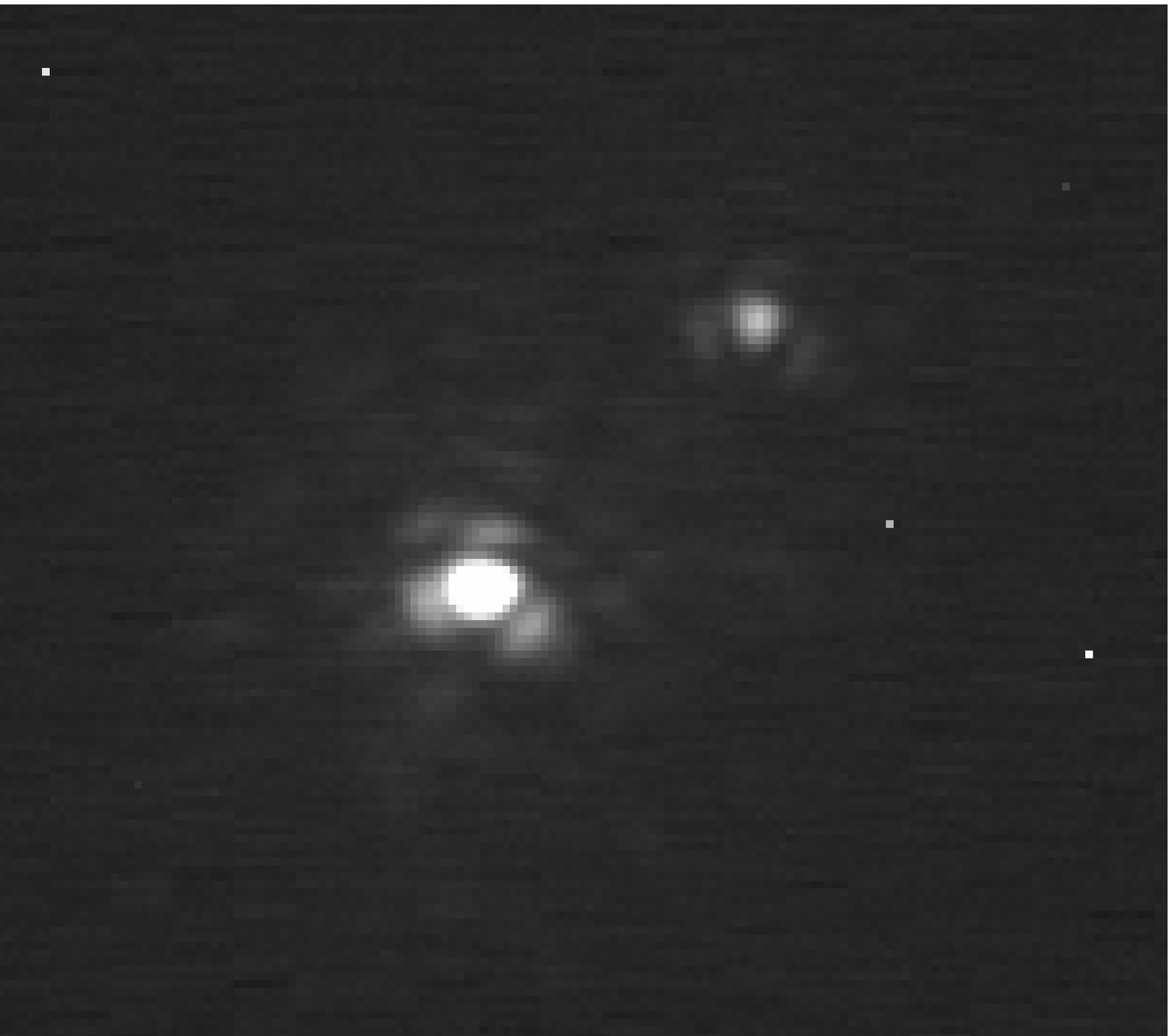}
  \end{minipage}
\caption{\label{fighii2147} {\it Top:} 
                Epoch 05 May 2012 VLBI image of the binary HII\,2147 system. Both 
                components are
                significantly detected. The projected separation between the
                binary components is 50\,milliarcseconds or roughly 6\,AU.
                {\it Bottom:} Three epochs of Keck~II NIRC2-NGSAO imaging of
                HII\,3197. The system is resolved into a triple and significant orbital motion
                of the close pair is seen between 2006 (left image), to 2011 (middle image),
                to 2012 (right image). North is up and East is left in each image. The separation
                between the close pair and the tertiary is $\approx$0.6$''$.}
\end{figure}


\begin{figure}
 \centering
 \includegraphics[width=110mm]{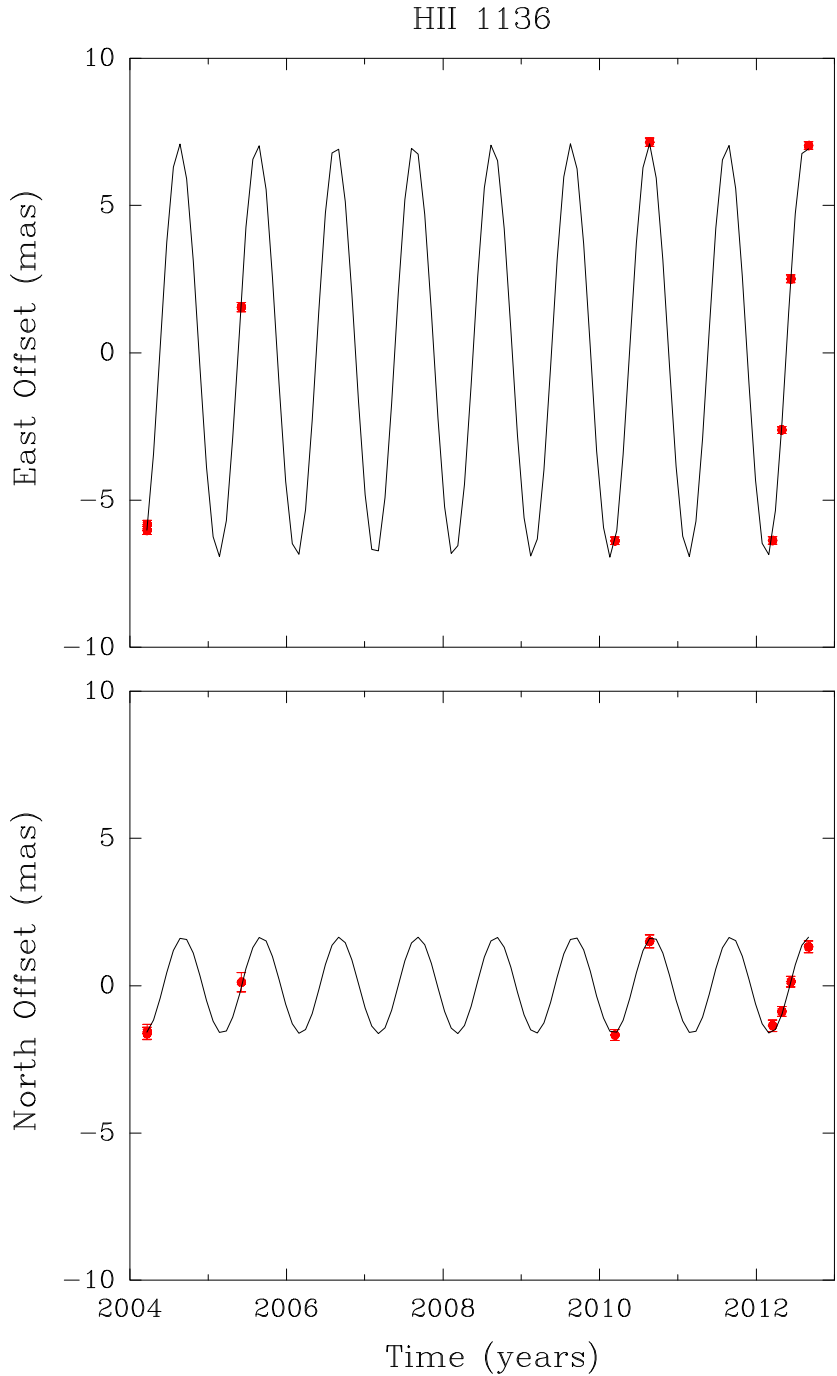}
\caption{\label{fighii1136} Astrometric model for Pleiad HII\,1136 that includes proper motion, 
               parallax, and orbital acceleration from an unseen binary companion. The top panel curve 
               and data points show right ascension angular offsets on the sky of the source position 
               relative to an arbitrary reference position. The bottom panel curve and data points show 
               declination offsets. Proper motion has been removed in the data points to accentuate the 
               parallax motion. The fit allows for acceleration as might come from a widely separated 
               (orbital period greater than 10 years) stellar companion. However, the fitted accelerations 
               are small and do not change the parallax.}
\end{figure}



\begin{discussion}
\end{discussion}

\end{document}

%% file: astmac.tex
\usepackage{color}
\usepackage{graphicx}

\newcommand{\apj}{ApJ}
\newcommand{\apjs}{ApJS}
\newcommand{\apjl}{ApJL}
\newcommand{\aap}{A\&A}

\newcommand{\aj}{AJ}
\newcommand{\nat}{Nature}